\documentclass[11pt]{article}

\usepackage[final]{acl}

\usepackage{times}
\usepackage{latexsym}

\usepackage[T1]{fontenc}

\usepackage[utf8]{inputenc}

\usepackage{microtype}

\usepackage{inconsolata}

\usepackage{graphicx}
\usepackage{tabularx}
\usepackage{array}
\usepackage{listings}
\usepackage{amssymb}
\usepackage{booktabs}
\usepackage{multirow}
\usepackage{pifont}
\usepackage{fontawesome5}
\usepackage{enumitem}
\usepackage[most]{tcolorbox}
\usepackage[table]{xcolor}

\usepackage{minted}
\setminted{
  fontsize=\footnotesize,
  breaklines,
  breakanywhere,
  autogobble,
  frame=single
}
\usepackage{svg}

\newcommand{\bench}{\textsc{ChainSWE}}
\newcommand{\modeSEQ}{\textsc{Seq}}
\newcommand{\modeMEM}{\textsc{Seq+Mem}}
\newcommand{\modeINT}{\textsc{Oracle}}
\newcommand{\sweedit}{\textsc{SWE-Edit}}
\newcommand{\agentSWE}{\textsc{SWE-agent}}

\newcommand{\cBase}{\textsc{Baseline}}
\newcommand{\cSum}{\textsc{Summarize}}
\newcommand{\cSub}{\textsc{Sub-Agent}}

\title{\bench{}: Benchmarking Coding Agents on Multi-Bug Software Maintenance} 

\author{
 \textbf{Qirui Jin\textsuperscript{1$\ddagger$}},
 \textbf{Lingching Tung\textsuperscript{4}},
 \textbf{Kenan Li\textsuperscript{2}},
 \textbf{Qiyang Shi\textsuperscript{3}},
 \textbf{Yushi She\textsuperscript{1}},
\\
 \textbf{Huanzhong Jia\textsuperscript{1}},
 \textbf{Harrison Zhao\textsuperscript{5}},
 \textbf{Kejing Xia\textsuperscript{1}},
 \textbf{Zhenbang Du\textsuperscript{1}},
 \textbf{Jiaxin Pei\textsuperscript{6}},
\\
 \textbf{Zhenyu Zhang\textsuperscript{6}},
 \textbf{Zhen Qi\textsuperscript{7}},
 \textbf{Yuyan Duan\textsuperscript{1}},
 \textbf{Wenke Lee\textsuperscript{1}},
 \textbf{Zijian Jin\textsuperscript{3$\ddagger$}}
\\
\\
 \textsuperscript{1}Georgia Institute of Technology \quad
 \textsuperscript{2}University College London \quad
 \textsuperscript{3}New York University
\\
 \textsuperscript{4}NVIDIA \quad
 \textsuperscript{5}Cornell University \quad
 \textsuperscript{6}Stanford University \quad
 \textsuperscript{7}Northeastern University
\\
 \small{
    $^\ddagger$\textbf{Correspondence:} \href{mailto:qiruijin@gatech.edu}{qiruijin@gatech.edu}, \href{mailto:zj2076@nyu.edu}{zj2076@nyu.edu}}
}

\begin{document}
\maketitle
\begin{abstract}
Language model (LM) agents are increasingly deployed to maintain codebases over extended periods, fixing streams of related defects while carrying context from one fix to the next.
Yet existing software engineering (SWE) benchmarks evaluate models one bug at a time: the repository is reset, the codebase is re-read, and a single self-contained issue is graded in isolation.
This setting collapses a continuous maintenance workflow into a series of independent sessions, ignoring the cumulative dependencies that make real-world bug fixing challenging.
To bridge this gap, we introduce \bench, the first benchmark for evaluating agents on sequential, dependent bug fixes within a shared codebase.
We collect chronological chains of $304$ issues across $54$ Python projects, mined from six SWE-bench-family datasets.
Our evaluation across a range of agents and models reveals a consistent performance drop by \textbf{up to 70\%} as the chain length increases.
Our data and code are available at \href{https://huggingface.co/datasets/C-lister/ChainSWE}{C-lister/ChainSWE}.
\end{abstract}
\section{Introduction}
\label{sec:intro}

Language model (LM) agents are now deployed across a wide range of software engineering tasks, including competitive programming~\citep{quan2025codeelo,jain2024livecodebench}, code repair~\citep{xia2024agentless,li2026oracle}, automatic environment setup~\citep{hu2026repo2run,li2026repolaunch}, and full-repository construction~\citep{ding2026nl2repo,yang2026programbenchlanguagemodelsrebuild}.
In particular, agent performance has improved rapidly on a family of issue-resolution benchmarks, including SWE-bench~\citep{jimenez2024swebench}, SWE-bench Live~\citep{zhang2025swebenchgoeslive}, SWE-rebench~\citep{badertdinov2025swerebenchautomatedpipelinetask} and SWE-rebench-v2~\citep{badertdinov2026swerebenchv2languageagnosticswe}, SWE-Gym~\citep{pan2025training}, and SWE-bench Pro~\citep{deng2025swe}.
Strong agents now successfully resolve a substantial fraction of issues on these live leaderboards~\citep{yang2024sweagent,wang2025openhands}.

However, every benchmark in this family follows the same basic evaluation protocol: a model is evaluated on a single, self-contained issue within a sandbox container.
The agent starts from a clean checkout at a pre-specified runnable base commit, receives a single problem statement, edits the codebase, and is evaluated against pre-written tests harvested from that issue.
This protocol measures whether a patch satisfies a local hard-coded test suite. However, it does not capture whether the agent avoids unnecessary edits, resolves latent multi-file dependencies, or manages context across an evolving sequence of related fixes.
As a result, this per-issue schema misses \textbf{two} failure modes that only emerge when bugs are evaluated sequentially, repository state carries across rollouts, and available tests fail to capture the damage left by earlier patches:

\begin{figure*}
    \centering
    \includegraphics[width=\linewidth]{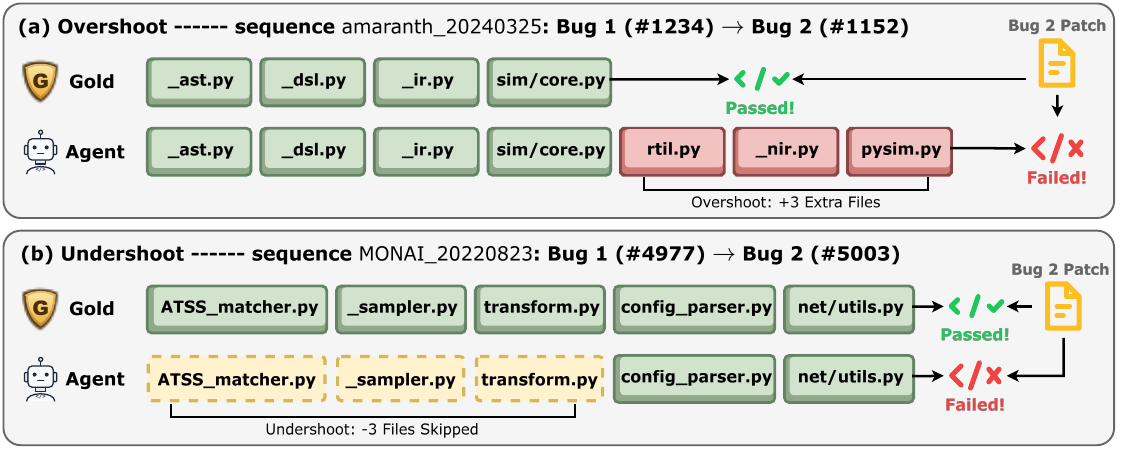}
    \caption{Two failure modes (\emph{overshoot} and \emph{undershoot}) exposed by \bench\ rollouts.}
    \label{fig:overshoot_undershoot}
\end{figure*}

\begin{itemize}[leftmargin=0.4cm]
\item \emph{Overshoot} (Figure~\ref{fig:overshoot_undershoot} (a)), 
where an agent fixing an earlier bug modifies files beyond the gold-patch-intended area; when a downstream bug subsequently targets those same files, its edits are applied on a polluted baseline, and otherwise correct changes no longer produce the expected behavior.
The \texttt{amaranth} chain in Figure~\ref{fig:overshoot_undershoot}(a) provides a representative example. The gold patch modifies four files, whereas the agent additionally rewrites three unrelated backend files. A subsequent issue targets two of these files, causing all eight downstream tests to fail.

\item \emph{Undershoot} (Figure~\ref{fig:overshoot_undershoot} (b)), where the prior bug's gold patch is a multi-file refactor, but the agent ships only the subset named in the issue title or description; downstream tests then fail not because the current bug is mishandled but because they exercise code paths whose preconditions were never established.
The \texttt{MONAI} chain in Figure~\ref{fig:overshoot_undershoot}(b) exhibits the opposite failure mode. Although the gold patch spans five files, the agent edits only two files explicitly referenced in the issue description and omits three supporting refactors. As a result, a downstream test fails despite the agent producing the correct modification to the target file.
\end{itemize}

These examples show a mismatch between current benchmarks and the real-world software maintenance process.
In practice, software engineers must leave the codebase in a state that is both correct for the current issue and usable as the starting point for future work.
They also need to read through the surrounding context and adapt their fixes to the evolving repository state they inherit.

To bridge this gap, we introduce \bench, a benchmark for \emph{continuous} SWE rollouts.
Each \bench{} instance consists of a time-ordered chain of real bug fixes with overlapping modified files, functions, or classes from the same repository.
We validate each chain by replaying the accumulated gold patches and associated test suites.
During evaluation, at step \(k\), the agent receives only the \(k\)-th issue statement and operates on the repository state produced by the previous steps.
The agent is evaluated not only on whether it can solve each issue in isolation, but also on whether its earlier patches create a valid substrate for later fixes.
We refer to failures caused by accumulated agent-generated state, rather than by the intrinsic difficulty of the current bug, as chain errors.
Our \bench{} dataset contains $100$ bug chains spanning $304$ instances and $54$ repositories, drawn from six open-source SWE-bench-family datasets.

We evaluate seven state-of-the-art language models using a fixed agent scaffold (\sweedit~\citep{zhang2026swe}) under three context-management configurations, where continuous rollout reveals failure modes that are not exposed under conventional isolated protocols.
We observe that performance drops by up to $70\%$ compared to the corresponding single-issue setting, with the sharpest declines at the deepest positions in a chain.
Our results suggest that further progress in SWE agents will require advances in dependency tracking, long-horizon reasoning, and robust repository-state management, rather than improvements in isolated issue resolution alone.

The contributions of this paper are threefold:
\begin{itemize}
  \item We introduce \bench, the first open-source SWE benchmark that explicitly evaluates language models on sequential, multi-bug rollouts over real-world repositories.
  \item We develop a scalable pipeline for constructing and validating chained SWE tasks from existing benchmarks, enabling continuous evaluation over arbitrarily long bug-fix sequences.
  \item We conduct comprehensive evaluations of seven state-of-the-art LMs under multiple context-management configurations.
  We provide detailed analysis of their failure modes, offering guidance for future work on dependency-aware reasoning in code LMs and repository state management in agent harnesses.
\end{itemize}
\section{Related Work}
\label{sec:related}
\paragraph{Repository-level issue resolution benchmarks.}
SWE-bench~\citep{jimenez2024swebench} introduced the standard setup of pairing real GitHub issues with hidden test suites to grade repository-level coding agents, and a number of variants have followed.
SWE-bench Verified re-audits and re-labels the original instances to remove under-specified or untestable tasks.
SWE-bench Live~\citep{zhang2025swebenchgoeslive}, SWE-rebench~\citep{badertdinov2025swerebenchautomatedpipelinetask} and SWE-rebench-v2~\citep{badertdinov2026swerebenchv2languageagnosticswe} propose automated pipelines that continuously curate up-to-date issues to mitigate training-set contamination, while SWE-Gym~\citep{pan2025training} targets training rather than evaluation and SWE-bench Pro~\citep{deng2025swe} extends both the difficulty and the size of the task set.
BeyondSWE~\citep{chen2026beyondswe} broadens code-agent evaluation beyond single-repository bug fixing to cross-repository, domain specific, dependency migration, and document-to-repository tasks.
ScaleSWE~\citep{zhao2026immersion} instead presents an automated pipeline for constructing large-scale, verified software engineering data for coding agent training.

Despite their differences, all of these benchmarks follow the same evaluation protocol: one issue per rollout, a fresh container reset to a clean base commit, and a fresh conversation, so each task is graded in complete isolation from every other.
This design measures \emph{single-shot} issue-resolution ability but deliberately removes the cross-task dependencies that characterize real maintenance, where a developer fixes a stream of related defects on top of a repository they themselves have just modified.
Unlike these benchmarks, \bench\ asks the agent to work through a chain of related issues in a single rollout, without resetting between bugs, and grades both per-bug and full-chain success to observe failures caused by accumulated, self-generated state.

\paragraph{Agent harnesses for code.}
A coding \emph{harness} connects a language model to tools—such as shell execution, file navigation, and code editing—and controls the context the model sees at each step.
Existing harnesses take different approaches to context management: \agentSWE~\citep{yang2024sweagent} uses a structured interface; \textsc{OpenHands}~\citep{wang2025openhands} automatically summarizes conversation history at context limit; \textsc{Claude Code}~\citep{claudecode} further proposes sub-agent delegation and task planning for better long-context management; and \textsc{Agentless}~\citep{xia2024agentless} replaces an autonomous loop with a fixed, step-by-step pipeline.
Related work also considers budget-aware observation planning and the reliability of self-improving agent workflows~\citep{peng2026bap,wang2026safe,liu2026operational,wang2026phantom}.
Most relevant to our study is \sweedit~\citep{zhang2026swe}, which uses a constrained edit format and an optional sub-agent to apply code changes, lowering token costs during long interactions.
These harnesses primarily differ in how they manage state across multiple turns.
We adopt \sweedit\ as a baseline framework to test three distinct context-management strategies: full history, summarization, and sub-agent editing.
Using \bench, we measure how each strategy performs as conversation history and code edits accumulate across a continuous, multi-bug session.

\paragraph{Sequential and continual reasoning benchmarks.}
Another line of work~\citep{liu2024agentbench, ma2024agentboard} evaluates language agents in multi-step environments where success requires planning, tool use, and state tracking.
Recent memory benchmarks further test whether assistants can retain and update information over long conversations or multi-session user histories~\citep{maharana2024evaluating, wu2025longmemeval}.
These settings are sequential, but they typically reset between tasks or evaluate memory over dialogue state rather than over persistent changes to an external artifact.
To the best of our knowledge, \bench\ is the first benchmark that combines real, test-validated repository bug fixes, multi-bug chains constructed with explicit code-overlap, and an evaluation protocol that distinguishes per-bug from full-chain success.
\section{\bench}
\label{sec:benchmark}

This section illustrates the data collection pipeline of \bench\ in detail.

\begin{figure*}[t]
    \centering
    \includegraphics[width=\linewidth]{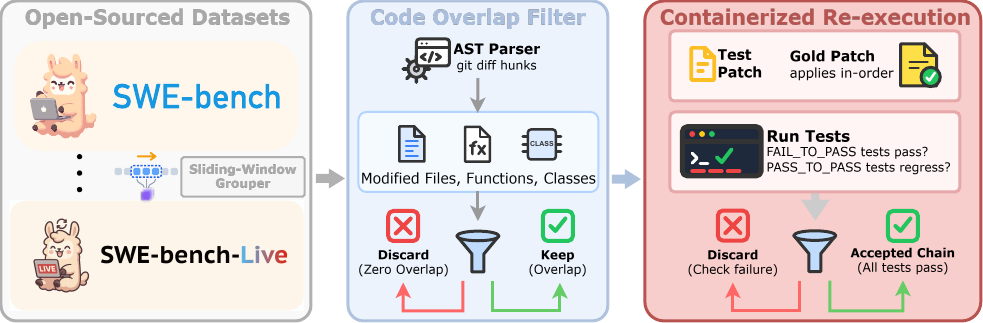}
    \caption{The data mining pipeline for \bench.}
    \label{fig:mining}
\end{figure*}

\subsection{Task Features}
\label{sec:benchmark:substrate}

\bench\ pools instances from six existing repository-level SWE
benchmarks: SWE-bench~\citep{jimenez2024swebench}, SWE-bench Live~\citep{zhang2025swebenchgoeslive}, SWE-rebench~\citep{badertdinov2025swerebenchautomatedpipelinetask} and SWE-rebench-v2~\citep{badertdinov2026swerebenchv2languageagnosticswe}, SWE-Gym~\citep{pan2025training}, and SWE-bench Pro~\citep{deng2025swe}.
Every instance is a tuple of repository, base commit, problem statement, gold patch, test patch, \texttt{FAIL\_TO\_PASS} test list, \texttt{PASS\_TO\_PASS} test list, and test commands.
All six benchmarks target Python projects and ship pre-built Docker images for reproducible test execution, which enables evaluation of an arbitrary subset under a single harness.
An instance of \bench\ is a unified list of such instances, each ordered chronologically with the Docker image reset to the base commit of the first bug in the chain, tagged with a unique chain ID.

\subsection{Mining Chains}
As illustrated in Figure~\ref{fig:mining}, we synthesize \bench\ chains in two stages: code-overlap scoring and containerized test re-execution.
For each repository in the pooled source, we collect all of its instances and sort them by commit date.
We then group consecutive instances into chains using a sliding window, joining two instances into the same chain if either of the following conditions holds:
\begin{enumerate}
    \item The latter's base commit is the immediate successor of the former's (i.e., no intervening commits in the repository's history).
    \item The latter's base commit is not the immediate successor of the former's, but its gold patch and test patch apply cleanly on top of the former's resolved state.
\end{enumerate}
The resulting candidate chains are passed to the filtering stages below; note that a chain of length $l$ also induces 
all contiguous sub-chains of lengths ranging from $2$ to $l$.

\textbf{Code-overlap heuristic filter.}
For each gold patch, we use an AST parser on the \texttt{git diff} hunks to extract the modified files, functions, and classes, and discard any window whose bugs have zero pairwise overlap.

\textbf{Test re-execution filter.}
The heuristic above is a cheap signal but does not guarantee that the chain actually builds.
We therefore run a containerized filter that, for each candidate chain, pulls the source benchmark's Docker image, starts a fresh container at the chain's base commit, and applies---in order---each bug's test patch followed by its gold patch (the same setting as \modeINT; see Section~\ref{sec:modes}).
Then, we run the bug's test commands and verify that all \texttt{FAIL\_TO\_PASS} tests now pass and no \texttt{PASS\_TO\_PASS} tests regress.
A chain is accepted only if all filter stages clear.

\subsection{Data Selection}
After running the pipeline above, we obtained candidate chains with lengths ranging from $2$ to $5$.
During data selection, we intentionally drop all chains of length 2 for two reasons:
\begin{enumerate}
    \item Length-$2$ chains are too short to constitute a meaningful stress test: with only one prior bug, there is no meaningful accumulation of context or compounding of patch errors.
    \item Most length-$2$ sequences are already embedded as contiguous prefixes inside length-3 or longer chains, so evaluating them separately would be largely redundant.
\end{enumerate}

For the remaining chains, we apply a second round of difficulty-based selection.
Because \bench\ targets the cross-task state-management capabilities of agent harnesses rather than raw bug-solving ability, each chain should be \emph{individually tractable} once state propagation is removed: a chain that fails because a single bug is intrinsically too hard tells us little about how well an agent carries state across fixes.
We therefore pre-run \modeINT\ mode (see Section~\ref{sec:modes}) on \cBase with \texttt{Qwen3.7-Max} and retain only chains for which more than half of the bugs are solved (e.g., for a chain of length $k$, at least $\lfloor k/2\rfloor+1$ bugs must pass).
This filters out chains whose difficulty is dominated by the per-bug task rather than by accumulated repository state.

\subsection{Dataset Statistics}
\label{sec:benchmark:stats}
The final version of \bench\ yields $100$ accepted chains with an average length of $3.04$, comprising $304$ bug-fix tasks drawn from $54$ unique repositories and spanning April 2019 to August 2025.
A detailed breakdown is provided in the Appendix~\ref{sec:data_detail}.
\section{Evaluation}

\subsection{Agent Configurations}
We hold the agent scaffold fixed across all experiments and vary only how it manages its context, so that performance differences can be attributed to the model and the context-management strategy rather than to the harness.
We use \sweedit~\citep{zhang2026swe}, a variant of \agentSWE; we evaluate three configurations that differ only in how the trajectory is kept within the context window as it grows:

\paragraph{\cBase.}
The base \sweedit\ agent with no augmentations, identical to the original \agentSWE.
It retains the full interaction transcript in context and applies edits with an exact string-replacement editor; no context compaction of any kind is performed.

\paragraph{\cSum.}
\cBase\ augmented with conversation summarization.
Once the running context exceeds a fixed budget ($50$K input tokens), we replace the earlier portion of the history with a model-written structured summary while retaining the $50$ most recent messages.

\paragraph{\cSub.}
\cSub\ offloads file viewing and editing to dedicated viewer and editor sub-agents, rather than placing file contents and diffs directly in the main agent's context.
The main agent issues a natural-language instruction describing the intended change, and a separate, typically smaller, editor model materializes the concrete patch.
This keeps the most token-heavy content out of the primary trajectory; no summarization is applied.

All three configurations share the same model, tools, and prompts (see Appendix~\ref{sec:agent_prompts}).

\subsection{Language Models}
We evaluate seven LMs spanning different long-context management capacities and tool-use strengths: \texttt{GPT-5.5}, \texttt{GPT-5.4-mini}, \texttt{GPT-5.4-nano}, \texttt{Claude-Opus-4.7}, \texttt{Claude-Opus\\-4.5}, \texttt{DeepSeek-V4-Pro}, \texttt{Gemini-3.1-Pro}.
All experiments use the API's default inference parameters.
For reasoning-capable models, we set thinking effort to \texttt{medium} to balance performance and cost.
Table~\ref{tab:model-contexts} summarizes the models used in this paper.

\begin{table}[t]
\centering
\small
\begin{tabular}{lll}
\toprule
\textbf{Model} & \textbf{SWE-bench Pro Score} & \textbf{Context}\\
\midrule
GPT-5.5 & $58.6\%$ & 1M \\
GPT-5.4-mini & $54.4\%$ & 400K \\
GPT-5.4-nano & $52.4\%$ & 400K \\
Claude-Opus-4.7 & $64.3\%$ & 1M \\
Claude-Opus-4.5 & $57.1\%$ & 200K \\
DeepSeek-V4-Pro & $54.4\%$ & 1M \\
Gemini-3.1-Pro & $54.2\%$ & 1M \\
\bottomrule
\end{tabular}
\caption{Language models evaluated on \bench. Exact API context limits are those exposed by the serving provider at evaluation time.}
\label{tab:model-contexts}
\end{table}

\subsection{Evaluation Modes}
\label{sec:modes}

\begin{figure}
    \centering
    \includegraphics[width=\linewidth]{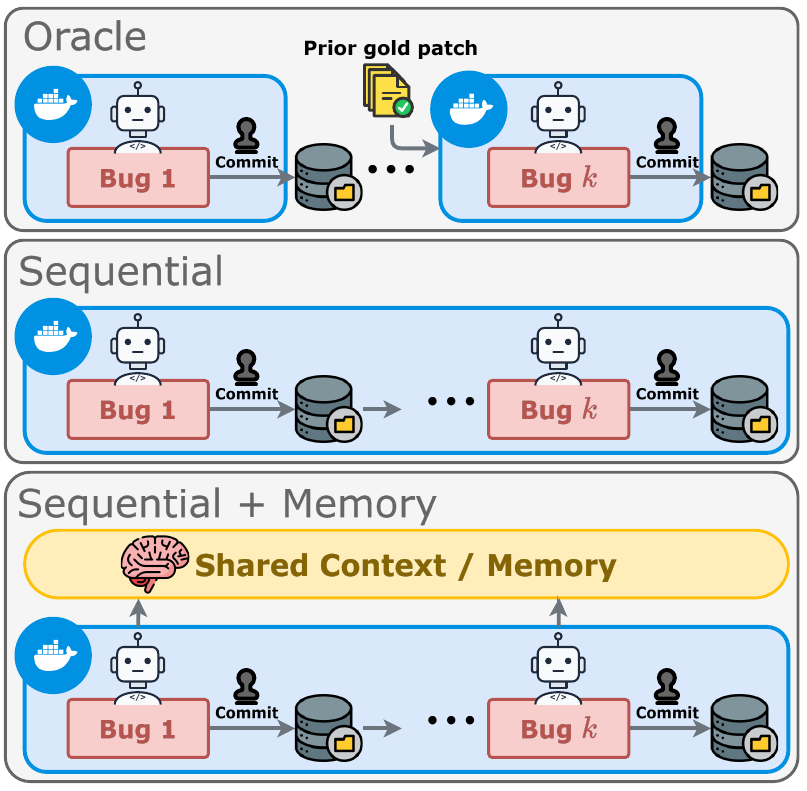}
    \caption{The three evaluation modes of \bench: \modeINT\ (oracle prior fixes), \modeSEQ\ (cumulative repository, fresh agent), and \modeMEM\ (cumulative repository, persistent agent).}
    \label{fig:modes}
\end{figure}

\bench\ scores each agent-model pair in three modes, as demonstrated in Figure~\ref{fig:modes}, each exercising a distinct dimension of context management.

\textbf{\modeINT}
is the chain-aware and self-contained analogue of the standard SWE-bench protocol. 
For each bug in the chain, the container is reset to the base commit, then all earlier bugs in the chain have their gold test patches and gold fix patches applied before the agent starts.
The agent never sees its own prior patches; instead, it operates on an oracle codebase matching the original maintenance history.

\textbf{\modeSEQ\ (Sequential).}
The agent works through bugs 1 to $n$ in order inside one container that is not reset between bugs.
After each bug, the model-generated patch is committed to the repository, and the next bug starts from this updated codebase.
The agent harness is restarted between bugs, so the conversation is fresh, but the repository carries the full history of model patches.
A chain is not aborted at the first or second failed bug to ensure a fair comparison with \modeINT\ because downstream fixes may still fix the chain cleanly.

\textbf{\modeMEM\ (Sequential + Memory).}
This mode is identical to \modeSEQ\ except that the agent's conversation history is also preserved across bugs.
The agent enters bug~$k{+}1$ with the transcript from bug~$k$, including intermediate observations and dead ends.
This mode isolates the effect of \emph{transcript-level memory} from \emph{repository-level state}.

\modeINT\ provides an oracle-state per-bug baseline, independent of error propagation, while \modeSEQ\ and \modeMEM\ measure the cost of carrying model-modified repository state across a chain.
The gap between these modes quantifies how much performance is lost due to accumulated agent and repo state rather than genuine bug difficulty.

\subsection{Experiment Settings}
All experiments are executed on a CPU-only Linux server (x86\_64) with 32 cores and 64\,GB of RAM, running 4 evaluations in parallel.
For all agents, we enforce a per-instance limit of 100 turns and a per-instance timeout of 30 minutes; all other parameters use the default configurations of each harness.

\subsection{Metrics}

We report \textbf{\%Resolved}, which refers to the percentage of task instances where the agent-generated patches pass all tests.
Because \bench\ evaluates ordered bug sequences, we also report per-chain success by position and full-chain success rate, where a chain is successful only if all bugs are resolved.
Finally, we report average API cost per task and per chain, computed from the input and output tokens consumed during each rollout.

\section{Results}

\newcommand{\drop}[1]{\,\textcolor{red}{\scriptsize $-$#1\%}}
\newcommand{\gain}[1]{\,\textcolor{green!55!black}{\scriptsize $+$#1\%}}

\begin{table*}[t]
  \centering
  \small
  \setlength{\tabcolsep}{3pt}
  \begin{tabular}{ll ccc ccc ccc}
    \toprule
    & & \multicolumn{3}{c}{\textbf{\cBase}} & \multicolumn{3}{c}{\textbf{\cSum}} & \multicolumn{3}{c}{\textbf{\cSub}} \\
    \cmidrule(lr){3-5} \cmidrule(lr){6-8} \cmidrule(lr){9-11}
    \textbf{Model} & \textbf{Mode}
      & \textbf{Per-Bug} & \textbf{Chain} & \textbf{Cost}
      & \textbf{Per-Bug} & \textbf{Chain} & \textbf{Cost}
      & \textbf{Per-Bug} & \textbf{Chain} & \textbf{Cost} \\
    & & (\%) & (\%) & (\$) & (\%) & (\%) & (\$) & (\%) & (\%) & (\$) \\
    \midrule
    \multirow{3}{*}{GPT-5.5}
      & \modeINT & $69.1$ & $26.0$ & $\phantom{0}6.0$
                  & $68.1$ & $23.0$ & $\phantom{0}6.2$
                  & $62.5$ & $22.0$ & $\phantom{0}4.8$ \\
      & \modeSEQ & $49.0$\drop{29} & $23.0$\drop{12} & $\phantom{0}5.6$
                  & $42.4$\drop{38} & $25.0$\gain{9} & $\phantom{0}6.2$
                  & $35.2$\drop{44} & $20.0$\drop{9} & $\phantom{0}4.6$ \\
      & \modeMEM & $47.4$\drop{31} & $19.0$\drop{27} & $\phantom{0}6.4$
                  & $49.7$\drop{27} & $23.0$\drop{0} & $\phantom{0}6.8$
                  & $43.4$\drop{31} & $21.0$\drop{5} & $\phantom{0}4.3$ \\
    \cmidrule(lr){1-11}
    \multirow{3}{*}{GPT-5.4-mini}
      & \modeINT & $61.8$ & $22.0$ & $\phantom{0}1.1$
                  & $61.8$ & $26.0$ & $\phantom{0}1.1$
                  & $36.2$ & $10.0$ & $\phantom{0}1.2$ \\
      & \modeSEQ & $41.8$\drop{32} & $15.0$\drop{32} & $\phantom{0}1.1$
                  & $39.8$\drop{36} & $21.0$\drop{19} & $\phantom{0}1.2$
                  & $21.7$\drop{40} & $10.0$\drop{0} & $\phantom{0}1.3$ \\
      & \modeMEM & $41.8$\drop{32} & $17.0$\drop{23} & $\phantom{0}1.4$
                  & $37.8$\drop{39} & $18.0$\drop{31} & $\phantom{0}1.2$
                  & $22.0$\drop{39} & $\phantom{0}9.0$\drop{10} & $\phantom{0}1.1$ \\
    \cmidrule(lr){1-11}
    \multirow{3}{*}{GPT-5.4-nano}
      & \modeINT & $64.8$ & $27.0$ & $\phantom{0}0.4$
                  & $62.8$ & $22.0$ & $\phantom{0}0.3$
                  & $37.8$ & $12.0$ & $\phantom{0}0.2$ \\
      & \modeSEQ & $45.1$\drop{30} & $21.0$\drop{22} & $\phantom{0}0.3$
                  & $33.9$\drop{46} & $16.0$\drop{27} & $\phantom{0}0.3$
                  & $23.0$\drop{39} & $12.0$\drop{0} & $\phantom{0}0.2$ \\
      & \modeMEM & $44.1$\drop{32} & $18.0$\drop{33} & $\phantom{0}0.4$
                  & $35.5$\drop{43} & $18.0$\drop{18} & $\phantom{0}0.4$
                  & $18.1$\drop{52} & $\phantom{0}9.0$\drop{25} & $\phantom{0}0.2$ \\
    \cmidrule(lr){1-11}
    \multirow{3}{*}{Claude-Opus-4.7}
      & \modeINT & $64.5$ & $22.0$ & $\phantom{0}1.8$
                  & $65.5$ & $24.0$ & $\phantom{0}1.8$
                  & $63.5$ & $21.0$ & $\phantom{0}3.2$ \\
      & \modeSEQ & $40.5$\drop{37} & $18.0$\drop{18} & $\phantom{0}1.7$
                  & $39.5$\drop{40} & $20.0$\drop{17} & $\phantom{0}1.6$
                  & $38.0$\drop{40} & $17.0$\drop{19} & $\phantom{0}3.1$ \\
      & \modeMEM & $39.5$\drop{39} & $17.0$\drop{23} & $\phantom{0}2.1$
                  & $41.5$\drop{37} & $20.0$\drop{17} & $\phantom{0}2.3$
                  & $40.0$\drop{37} & $19.0$\drop{10} & $\phantom{0}3.6$ \\
    \cmidrule(lr){1-11}
    \multirow{3}{*}{Claude-Opus-4.5}
      & \modeINT & $64.0$ & $21.0$ & $\phantom{0}1.5$
                  & $66.5$ & $24.0$ & $\phantom{0}1.5$
                  & $62.5$ & $20.0$ & $\phantom{0}5.6$ \\
      & \modeSEQ & $43.0$\drop{33} & $19.0$\drop{10} & $\phantom{0}1.5$
                  & $40.5$\drop{39} & $18.0$\drop{25} & $\phantom{0}1.4$
                  & $38.5$\drop{38} & $17.0$\drop{15} & $\phantom{0}5.2$ \\
      & \modeMEM & $37.5$\drop{41} & $16.0$\drop{24} & $\phantom{0}3.1$
                  & $40.0$\drop{40} & $20.0$\drop{17} & $\phantom{0}2.9$
                  & $40.0$\drop{36} & $20.0$\drop{0} & $\phantom{0}6.7$ \\
    \cmidrule(lr){1-11}
    \multirow{3}{*}{DeepSeek-V4-Pro}
      & \modeINT & $42.8$ & $10.0$ & $\phantom{0}1.5$
                  & $50.3$ & $16.0$ & $\phantom{0}1.9$
                  & $48.4$ & $12.0$ & $\phantom{0}2.0$ \\
      & \modeSEQ & $29.6$\drop{31} & $\phantom{0}5.0$\drop{50} & $\phantom{0}1.5$
                  & $27.3$\drop{46} & $11.0$\drop{31} & $\phantom{0}1.9$
                  & $25.0$\drop{48} & $11.0$\drop{8} & $\phantom{0}2.0$ \\
      & \modeMEM & $28.6$\drop{33} & $\phantom{0}9.0$\drop{10} & $\phantom{0}1.7$
                  & $26.0$\drop{48} & $\phantom{0}8.0$\drop{50} & $\phantom{0}2.0$
                  & $28.0$\drop{42} & $11.0$\drop{8} & $\phantom{0}2.1$ \\
    \cmidrule(lr){1-11}
    \multirow{3}{*}{Gemini-3.1-Pro}
      & \modeINT & $61.8$ & $20.0$ & $\phantom{0}1.3$
                  & $61.5$ & $19.0$ & $\phantom{0}1.5$
                  & $61.5$ & $20.0$ & $\phantom{0}1.9$ \\
      & \modeSEQ & $36.5$\drop{41} & $18.0$\drop{10} & $\phantom{0}1.4$
                  & $39.5$\drop{36} & $20.0$\gain{5} & $\phantom{0}1.4$
                  & $37.2$\drop{40} & $19.0$\drop{5} & $\phantom{0}1.9$ \\
      & \modeMEM & $37.5$\drop{39} & $18.0$\drop{10} & $\phantom{0}1.9$
                  & $39.8$\drop{35} & $21.0$\gain{11} & $\phantom{0}1.9$
                  & $37.2$\drop{40} & $19.0$\drop{5} & $\phantom{0}2.5$ \\
    \bottomrule
  \end{tabular}
  \caption{Main results on \bench\ with \sweedit\ under three context-management configurations (\cBase, \cSum, \cSub).
  We report all $100$ chains ($304$ bugs); under \modeSEQ\ and \modeMEM.
  \textbf{Per-Bug} is per-bug accuracy out of the $304$ bugs; \textbf{Chain} is the fraction of the $100$ chains for which all bugs pass; \textbf{Cost} is the estimated average USD per chain.
  Red subscripts on \modeSEQ\ and \modeMEM\ rows give the relative change from the corresponding \modeINT\ baseline (green $+$ marks an increase).}
  \label{tab:main}
\end{table*}

\subsection{Overall Success Rate}

Because \modeINT\ already presents the agent with every earlier issue and its gold patch, the gap between \modeINT\ and the two sequential modes measures the cost of operating on imperfect, self-generated repository state.
The remaining gap between \modeSEQ\ and \modeMEM\ measures the effect of carrying the interaction history across bugs while still solving on self-generated state.

\paragraph{Sequential execution sharply reduces per-bug accuracy.}
Averaged over the seven models and three configurations, per-bug accuracy drops from $58.9\%$ under \modeINT\ to $36.5\%$ under \modeSEQ\ and $36.9\%$ under \modeMEM, a roughly $38\%$ relative decline.
Every one of the $21$ model$\times$configuration cells loses accuracy when moving from \modeINT\ to \modeSEQ, so the effect is not driven by a few weak models or settings.
For \modeSEQ, the decline ranges from $29\%$ for \texttt{GPT-5.5} under \cBase\ ($69.1 \to 49.0$) to $48\%$ for \texttt{DeepSeek-V4-Pro} under \cSub\ ($48.4 \to 25.0$).
Chain accuracy is less diagnostic because a single missed bug makes the whole chain fail; it is already low under \modeINT\ (mean $20.0\%$).
Under \modeSEQ\ the mean is comparable ($17.0\%$), most model--configuration pairs fall, as later bugs inherit the state drift and errors left behind by earlier self-generated edits.
A few actually improve, e.g.\ \texttt{GPT-5.5} and \texttt{Gemini-3.1-Pro} under \cSum.
This is because a correct earlier fix leaves behind context that helps with the next bug.

\paragraph{Robustness to accumulated state varies across models.}
The relative \modeINT-to-\modeSEQ\ degradation in per-bug accuracy, averaged over configurations, ranges from $36\%$ for \texttt{GPT-5.4-mini} to $42\%$ for \texttt{DeepSeek-V4-Pro}.
The most robust models are those able to tell which of the previously accumulated edits are relevant to the current bug and which are not, and to reason about the effects those edits have already had on the repository.
Under the raw-history baseline \cBase, \texttt{GPT-5.5} is the most robust model ($-29\%$), but its advantage erodes under the rewriting configurations ($-38\%$ under \cSum, $-44\%$ under \cSub).
In comparison, smaller GPT models show more consistent and uniform degradation.
Among the models with competitive \modeINT\ accuracy, \texttt{Claude-Opus-4.7} shows one of the largest average gaps ($-39\%$), indicating that strong single-shot ability does not guarantee robustness to self-generated state.

\paragraph{\texttt{GPT-5.5} shows the largest observed gain from conversation memory.}
Carrying the conversation history across bugs (\modeMEM) does not improve the performance, as mean per-bug accuracy is essentially unchanged from \modeSEQ\ ($36.9\%$ vs.\ $36.5\%$).
\texttt{GPT-5.5} improves from a config-averaged $42.2\%$ under \modeSEQ\ to $46.8\%$ under \modeMEM, with gains of $+7.3$ points under \cSum\ ($42.4 \to 49.7$) and $+8.2$ points under \cSub\ ($35.2 \to 43.4$).
The other six models see no such benefit: \modeMEM\ stays within about two points of \modeSEQ\ on average and is often slightly worse, with even \texttt{DeepSeek-V4-Pro} essentially unchanged (config-averaged $27.3 \to 27.5$).
Preserved history compensates for imperfect repository state only when the model can productively reason over the longer history; otherwise, it adds input tokens without a measurable benefit.

\paragraph{Lighter context management is more robust than active rewriting.}
Holding the model fixed, the context-management configuration determines how quickly sequential execution degrades.
Averaged over models, the relative \modeINT-to-\modeSEQ\ drop grows from $33\%$ under \cBase\ to $40\%$ under \cSum\ and $41\%$ under \cSub.
Simply retaining the raw history (\cBase) is therefore more robust than summarizing it (\cSum) or routing edits through an LLM editor (\cSub). Even though the latter two are designed to reduce context pressure, they degrade overall performance due to potential information loss and inconsistency.

\paragraph{Cost.}
Per-chain cost is driven primarily by model tier rather than execution mode: it ranges from $\$0.2$--$\$0.4$ for \texttt{GPT-5.4-nano} to $\$4.3$--$\$6.8$ for \texttt{GPT-5.5}.
Within a model, \modeMEM\ is often slightly more expensive than \modeSEQ\ because it carries longer histories, but the added context rarely translates into higher accuracy.
\cSum\ and \cSub\ can sometimes reduce cost for expensive main-model runs by shortening or moving token-heavy context out of the main trajectory, but Table~\ref{tab:main} shows that this saving often comes with lower robustness.

\begin{table}[t]
  \centering\small
  \setlength{\tabcolsep}{2pt}
  \begin{tabular}{llccc}
    \toprule
    \textbf{Context} & \textbf{Mode} &
      \textbf{Pos.\,1} & \textbf{Pos.\,2} & \textbf{Pos.\,3} \\
    \midrule
    \multirow{3}{*}{\cBase}
      & $\modeINT$ & $58.6$ & $64.3$ & $67.0$ \\
      & $\modeSEQ$ & $57.1$ & $39.3$\drop{39} & $27.7$\drop{59} \\
      & $\modeMEM$ & $56.5$ & $38.0$\drop{41} & $28.2$\drop{58} \\
    \cmidrule(lr){1-5}
    \multirow{3}{*}{\cSum}
      & $\modeINT$ & $58.0$ & $66.0$ & $68.4$ \\
      & $\modeSEQ$ & $60.0$ & $35.4$\drop{46} & $20.6$\drop{70} \\
      & $\modeMEM$ & $59.3$ & $36.6$\drop{45} & $22.0$\drop{68} \\
    \cmidrule(lr){1-5}
    \multirow{3}{*}{\cSub}
      & $\modeINT$ & $49.5$ & $58.1$ & $59.2$ \\
      & $\modeSEQ$ & $50.3$ & $29.9$\drop{49} & $17.5$\drop{70} \\
      & $\modeMEM$ & $50.4$ & $31.3$\drop{46} & $19.2$\drop{67} \\
    \bottomrule
  \end{tabular}
  \caption{Per-position resolution rate (\%) by context-management configuration, averaged across the seven evaluated models. Red subscripts report the relative drop from the corresponding \modeINT\ baseline.}
  \label{tab:position_config}
\end{table}

\begin{table}[t]
  \centering\small
  \setlength{\tabcolsep}{3pt}
  \renewcommand{\arraystretch}{1.05}
  \begin{tabular}{lccc}
    \toprule
    \textbf{Mode} & \textbf{Pos.\,1} & \textbf{Pos.\,2} & \textbf{Pos.\,3} \\
    \midrule
    \rowcolor{gray!20}\textbf{GPT-5.5}&&&\\
    $\modeINT$ & $61.2$ & $70.4$ & $75.9$ \\
    $\modeSEQ$ & $62.5$ & $40.5$\drop{42} & $28.2$\drop{63} \\
    $\modeMEM$ & $61.9$ & $44.7$\drop{37} & $37.8$\drop{50} \\
    \cmidrule(lr){1-4}
    \rowcolor{gray!20}\textbf{GPT-5.4-mini}&&&\\
    $\modeINT$ & $51.2$ & $58.1$ & $56.4$ \\
    $\modeSEQ$ & $51.9$ & $33.7$\drop{42} & $20.6$\drop{63} \\
    $\modeMEM$ & $51.5$ & $31.6$\drop{46} & $21.3$\drop{62} \\
    \cmidrule(lr){1-4}
    \rowcolor{gray!20}\textbf{GPT-5.4-nano}&&&\\
    $\modeINT$ & $50.2$ & $61.9$ & $59.1$ \\
    $\modeSEQ$ & $50.5$ & $32.0$\drop{48} & $21.6$\drop{63} \\
    $\modeMEM$ & $48.8$ & $30.9$\drop{50} & $21.0$\drop{65} \\
    \cmidrule(lr){1-4}
    \rowcolor{gray!20}\textbf{Claude-Opus-4.7}&&&\\
    $\modeINT$ & $60.0$ & $68.0$ & $72.0$ \\
    $\modeSEQ$ & $60.0$ & $38.0$\drop{44} & $24.0$\drop{67} \\
    $\modeMEM$ & $60.5$ & $39.0$\drop{43} & $25.0$\drop{65} \\
    \cmidrule(lr){1-4}
    \rowcolor{gray!20}\textbf{Claude-Opus-4.5}&&&\\
    $\modeINT$ & $60.5$ & $68.5$ & $71.0$ \\
    $\modeSEQ$ & $61.0$ & $39.0$\drop{43} & $24.5$\drop{65} \\
    $\modeMEM$ & $60.0$ & $38.0$\drop{45} & $23.0$\drop{68} \\
    \cmidrule(lr){1-4}
    \rowcolor{gray!20}\textbf{DeepSeek-V4-Pro}&&&\\
    $\modeINT$ & $44.3$ & $48.8$ & $53.3$ \\
    $\modeSEQ$ & $44.7$ & $24.7$\drop{49} & $15.1$\drop{72} \\
    $\modeMEM$ & $45.4$ & $26.1$\drop{46} & $14.1$\drop{74} \\
    \cmidrule(lr){1-4}
    \rowcolor{gray!20}\textbf{Gemini-3.1-Pro}&&&\\
    $\modeINT$ & $60.1$ & $63.9$ & $66.3$ \\
    $\modeSEQ$ & $59.8$ & $36.1$\drop{44} & $19.6$\drop{70} \\
    $\modeMEM$ & $59.8$ & $36.8$\drop{42} & $19.9$\drop{70} \\
    \bottomrule
  \end{tabular}
  \caption{Per-position resolution rate (\%) by model, averaged across the three context-management configurations. Red subscripts report the relative drop from the corresponding \modeINT\ baseline.}
  \label{tab:position_model}
\end{table}

\subsection{Success Rate by Position}

Tables~\ref{tab:position_config} and~\ref{tab:position_model} break per-bug resolution down by a bug's position in its chain, computed over the $97$ length-$3$ chains; the $3$ longer chains (two of length~$4$, one of length~$5$) are too few for stable per-position estimates and are omitted.
Position resolves the ambiguity in the aggregate numbers: it separates whether later bugs are intrinsically harder from whether accumulated state makes them harder.

\paragraph{Under oracle resets, later bugs are not harder.}
In \modeINT, each bug begins from the correct prior state, so by construction position reflects only intrinsic task difficulty rather than accumulated error; resolution is therefore expected to be flat or even rising with position.
Averaged over configurations, \texttt{GPT-5.5} resolves $61.2\%$, $70.4\%$, and $75.9\%$ at positions $1$ to $3$ (rising with position), and the same shape holds across
models (\texttt{Claude-Opus-4.7}: $60.0/68.0/72.0$; \texttt{DeepSeek-V4-Pro}: $44.3/48.8/53.3$).
Later bugs in a chain are therefore not inherently more difficult.

\paragraph{Under accumulation, difficulty compounds with depth.}
In both \modeSEQ\ and \modeMEM, an agent's success rate drops sharply and consistently along the chain. Under \modeSEQ, \texttt{GPT-5.5} falls from $62.5\%$ at position~$1$ to $40.5\%$ and then $28.2\%$; \texttt{GPT-5.4-nano} similarly drops from $50.5\%$ to $32.0\%$ to $21.6\%$.
For the first-bug success rate across all seven models, it remains close to the isolated \modeINT\ baseline (e.g., $62.5\%$ vs.\ $61.2\%$ for \texttt{GPT-5.5}), because the agent faces a clean slate with no accumulated changes.
The entire penalty for sequential fixing is therefore concentrated at positions $2$ and $3$.
As the chain gets deeper, the performance gap between isolated (\modeINT) and sequential (\modeSEQ) fixing widens significantly---eventually reaching a $37.5$-point drop for \texttt{GPT-5.4-nano} at position~$3$ ($59.1\%$ down to $21.6\%$).

\paragraph{The configuration ordering is stable across positions.} 
Table~\ref{tab:position_config} shows that the configurations are roughly the same at position~$1$ ($50$--$60\%$ across all modes) and separate only as state accumulates.
At position~$3$ under \modeSEQ, \cBase\ retains $27.7\%$ against $20.6\%$ for \cSum\ and $17.5\%$ for \cSub---a $59\%$ relative drop from the \modeINT\ baseline for \cBase\ versus $70\%$ for the rewriting configurations.
\modeMEM\ provides a small, inconsistent improvement at the last position (e.g., \cSub, $17.5 \to 19.2$) but does not change the ordering.
This is consistent with the model-level finding: under accumulation, context rewriting increases errors more than carrying the context in its original form.

\begin{figure}[t]
  \centering
  \includegraphics[width=\linewidth]{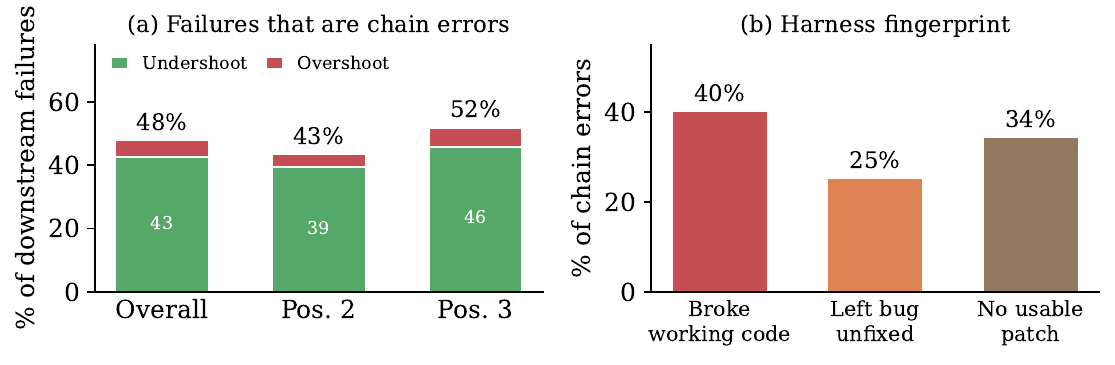}
  \caption{Chain-error analysis on \cBase\ cells. \textbf{(a)} Of the downstream \modeSEQ\ failures (positions~$2$--$3$), $48\%$ overall ($43\%$ at position~$2$, $52\%$ at position~$3$) are \emph{chain errors}; each bar is split into the share of chain errors caused by an earlier \emph{under-edit} (green) versus \emph{over-edit} (red), roughly nine to one. \textbf{(b)} Harness fingerprint of the $318$ chain errors that the agent attempted on the accumulated state.}
  \label{fig:chain_errors}
\end{figure}

\subsection{Failure Mode Analysis}
\label{sec:failure_modes}
Since agents struggle increasingly with later bugs, we shift our focus to analyze failures at the chain level rather than the individual bug level, where the latter has already been thoroughly explored in prior works~\citep{jimenez2024swebench,zhang2025swebenchgoeslive}.
For every downstream bug (position~$\geq 2$), we re-tested it in isolation (\modeINT) starting from a \emph{perfect} prior state. 
If a bug passes under \modeINT\ but fails under \modeSEQ, the failure is clearly induced by the agent's accumulated errors, not the bug itself. 
We call such a failure a \emph{chain error}.
Across the \cBase\ models, $318$ of the $663$ downstream \modeSEQ\ failures at positions~$2$--$3$ ($48\%$) are chain errors, and the rate grows with depth ($43\%$ at position~$2$ and $52\%$ at position~$3$; Figure~\ref{fig:chain_errors}a): once state accumulates, roughly half of all later-bug failures are attributable to the chain rather than the bug itself.

As introduced in Section~\ref{sec:intro}, depending on how an earlier bug's agent patch is miscalibrated \emph{relative to that earlier bug's own gold patch}, chain errors divide into two symmetric families: \textbf{Overshoot} and \textbf{Undershoot}. 
These two errors aren't mutually exclusive---by the time an agent reaches the third bug, earlier patches have typically both over- and under-edited the codebase.
However, when we can trace a chain error back to a single shared file, prior omissions (under-edits) clearly outnumber prior over-edits (Figure~\ref{fig:chain_errors}a).
This mirrors our earlier finding: agents are much more likely to leave a later bug's preconditions unmet than to actively overwrite working code.

Regardless of the cause, the chain errors the agent attempts on the accumulated state show up in the testing harness in one of three distinct ways.
As Figure~\ref{fig:chain_errors}b shows, the most common outcome is that the agent breaks previously working code simply because it is forced to build upon an already corrupted repository state:
\begin{itemize}
  \item \textbf{Broke working code} ($40\%$): the later bug's edit composes onto the corrupted state and a previously-passing \textsc{pass\_to\_pass} test breaks.
  \item \textbf{Left the bug unfixed} ($25\%$): the later bug's own \textsc{fail\_to\_pass} target cannot be made to pass because a precondition is missing, but nothing else breaks.
  \item \textbf{Produced no usable patch} ($34\%$): the inherited state prevents the agent from producing or applying a usable patch at all, or leaves the suite uncollectable.
\end{itemize}
The fact that \textbf{broke working code} is the most common outcome---around $40\%$ at both downstream positions---helps explain why preserving a richer chat history (\modeMEM) rarely helps.
The structural damage is already baked into the repository state before the agent even begins tackling the later bug, meaning a longer conversation history simply cannot undo it.

\section{Conclusion}
We introduced \bench, the first benchmark that evaluates coding agents on sequential, dependent bug fixes within a shared codebase.
By mining chains of real issues from six SWE-bench-family datasets and validating them through containerized test replay, \bench\ exposes failure modes that are not exposed when bugs are graded in isolation, with per-bug success dropping by up to $70\%$ at the deepest chain positions.
Over models and context-management configurations, we demonstrate that carrying the history of previous fixes yields only a marginal gain, whereas active summarization and sub-agent editing exhibit larger degradation.
These results offer insights for future work on long-context, dependency-aware reasoning in code LMs and on repository-state management in agent harnesses.

\subsection*{Limitations}
The six source benchmarks used to construct chain tasks inevitably contain noisy instances, where tests enforce implementation details or extra functionality not specified in the problem description~\cite{openai2026swebenchverified}. 
As a result, a small portion of chain failures can stem not from agents incorrectly following earlier issue descriptions, but from the impossibility of inferring the hidden requirements of later tests from the problem descriptions of earlier issues.

The current release is restricted to Python and is modest in scale.
Because long contiguous issue sequences are rare in existing benchmarks, we are unable to construct chain tasks longer than five instances from the current data pool.
This limits both the generalizability of our findings to other
programming languages and ecosystems and our ability to evaluate error
accumulation and context management over longer maintenance horizons.
Nonetheless, the construction pipeline itself is not specific to Python and can be extended as executable SWE-style datasets expand in language coverage, repository coverage, and size.

\subsection*{Acknowledgements}
The views expressed in this paper are solely those of the authors and do not reflect those of NVIDIA, which does not sponsor or endorse this research.

\bibliography{custom}

\appendix
\section{Dataset Details}
\label{sec:data_detail}

\subsection{Data Format}
\label{sec:data_format}

Each \bench\ instance is a \emph{chain}: a time-ordered list of bug-fix tasks over a single repository, sharing one Docker image and base commit.
Table~\ref{tab:chain-fields} lists the chain-level fields, and Table~\ref{tab:bug-fields} the per-bug fields stored inside \texttt{bug\_fixes}.
Each per-bug entry is itself a self-contained, SWE-bench-style task instance.

\begin{table*}[t]
\centering
\caption{Chain-level fields of a \bench\ instance.}
\label{tab:chain-fields}
\small
\begin{tabular}{lp{2.6cm}p{8.6cm}}
\toprule
\textbf{Field} & \textbf{Type} & \textbf{Description} \\
\midrule
\texttt{continuous\_id}        & \texttt{str}       & Unique identifier for the chain. \\
\texttt{repo}                  & \texttt{str}       & Source repository in \texttt{owner/name} form. \\
\texttt{base\_commit}          & \texttt{str}       & Commit the chain starts from; the repository is reset here before the first bug. \\
\texttt{source}                & \texttt{str}       & Originating SWE-bench-family dataset. \\
\texttt{date\_range}           & \texttt{str}       & Span of the earliest to latest bug date in the chain. \\
\texttt{docker\_image}         & \texttt{str}       & Pre-built execution environment shared by every bug in the chain. \\
\texttt{swebench\_instance\_ids} & \texttt{List[str]} & Ordered list of the underlying SWE-bench instance IDs in the chain. \\
\texttt{bug\_fixes}            & \texttt{List[dict]} & The ordered bug-fix tasks; each entry has the fields in Table~\ref{tab:bug-fields}. \\
\bottomrule
\end{tabular}
\end{table*}

\begin{table*}[t]
\centering
\caption{Per-bug fields stored in each entry of \texttt{bug\_fixes}.}
\label{tab:bug-fields}
\small
\begin{tabular}{lp{2.6cm}p{8.6cm}}
\toprule
\textbf{Field} & \textbf{Type} & \textbf{Description} \\
\midrule
\texttt{order}                & \texttt{int}       & Position of the bug within the chain ($1$-indexed). \\
\texttt{swebench\_instance\_id} & \texttt{str}     & Identifier of the underlying SWE-bench task instance. \\
\texttt{commit\_hash}         & \texttt{str}       & Commit at which this bug's gold patch was merged. \\
\texttt{problem\_statement}   & \texttt{str}       & Issue description used as the task prompt for this bug. \\
\texttt{date}                 & \texttt{str}       & Date of the bug's fixing commit, used to order the chain. \\
\texttt{source}               & \texttt{str}       & Originating SWE-bench-family dataset for this bug. \\
\texttt{patch}                & \texttt{str}       & Gold patch proposed by the pull request. \\
\texttt{test\_patch}          & \texttt{str}       & Modifications to the test suite to check whether the issue is resolved. \\
\texttt{FAIL\_TO\_PASS}       & \texttt{List[str]} & Tests expected to change from failing to passing once fixed. \\
\texttt{PASS\_TO\_PASS}       & \texttt{List[str]} & Tests already passing before the fix. \\
\texttt{test\_cmds}           & \texttt{str}       & Command(s) used to run the test suite. \\
\bottomrule
\end{tabular}
\end{table*}

\begin{table}[t]
\centering
\caption{Per-Source breakdown for \bench.}
\label{tab:src-breakdown}
\begin{tabular}{lc}
\toprule
Source & Chains \\
\midrule
SWE-rebench-v2     & $52$ \\
SWE-rebench        & $17$ \\
SWE-bench Live     & $12$ \\
SWE-Gym            & $\phantom{0}7$ \\
SWE-bench          & $\phantom{0}6$ \\
Mixed              & $\phantom{0}6$ \\
SWE-bench Pro      & $\phantom{0}0$ \\
\bottomrule
\end{tabular}
\end{table}

\begin{table}[t]
\centering
\caption{Average maximum context reached per chain, averaged across all models.}
\label{tab:context-usage}
\small
\begin{tabular}{lccc}
\toprule
\textbf{Config} & \modeINT & \modeSEQ & \modeMEM \\
\midrule
\cBase & $48.2$k & $48.1$k & $63.6$k \\
\cSum  & $42.7$k & $41.8$k & $48.4$k \\
\cSub  & $42.8$k & $42.6$k & $53.0$k \\
\bottomrule
\end{tabular}
\end{table}

\subsection{Per-Repository Breakdown}

\bench\ spans $54$ repositories with varying chain counts.
Table~\ref{tab:repo-breakdown} lists all $54$ repositories with their chain and bug counts; \texttt{tobymao/sqlglot} ($12$ chains) and \texttt{s-knibbs/dataclasses-jsonschema} ($7$) are the largest.

\begin{table*}[t]
\centering
\caption{Per-repository breakdown of \bench\ ($54$ repositories, $100$ chains, $304$ bugs), sorted by chain count.}
\label{tab:repo-breakdown}
\small
\begin{tabular}{lcc@{\hskip 2em}lcc}
  \toprule
  \textbf{Repository} & \textbf{Chains} & \textbf{Bugs} & \textbf{Repository} & \textbf{Chains} & \textbf{Bugs} \\
  \midrule
  \texttt{tobymao/sqlglot} & 12 & 36 & \texttt{drivendataorg/cloudpathlib} & 1 & 3 \\
  \texttt{s-knibbs/dataclasses-jsonschema} & 7 & 21 & \texttt{falconry/falcon} & 1 & 4 \\
  \texttt{Project-MONAI/MONAI} & 4 & 12 & \texttt{fonttools/fonttools} & 1 & 3 \\
  \texttt{brazilian-utils/brutils-python} & 4 & 12 & \texttt{hdmf-dev/hdmf} & 1 & 3 \\
  \texttt{pyccel/pyccel} & 4 & 12 & \texttt{hikari-py/hikari} & 1 & 3 \\
  \texttt{reata/sqllineage} & 4 & 12 & \texttt{instructlab/instructlab} & 1 & 3 \\
  \texttt{HypothesisWorks/hypothesis} & 3 & 9 & \texttt{ipython/ipython} & 1 & 3 \\
  \texttt{MartinThoma/flake8-simplify} & 3 & 9 & \texttt{jazzband/tablib} & 1 & 3 \\
  \texttt{PyCQA/docformatter} & 3 & 9 & \texttt{joke2k/faker} & 1 & 3 \\
  \texttt{amaranth-lang/amaranth} & 3 & 9 & \texttt{jsonpickle/jsonpickle} & 1 & 3 \\
  \texttt{pydata/xarray} & 3 & 9 & \texttt{keras-team/keras} & 1 & 3 \\
  \texttt{sdv-dev/RDT} & 3 & 9 & \texttt{koxudaxi/datamodel-code-generator} & 1 & 3 \\
  \texttt{getmoto/moto} & 2 & 6 & \texttt{matthewwithanm/python-markdownify} & 1 & 3 \\
  \texttt{litestar-org/polyfactory} & 2 & 6 & \texttt{meltano/sdk} & 1 & 3 \\
  \texttt{mikedh/trimesh} & 2 & 7 & \texttt{oscal-compass/compliance-trestle} & 1 & 3 \\
  \texttt{modin-project/modin} & 2 & 6 & \texttt{pallets/click} & 1 & 3 \\
  \texttt{mwaskom/seaborn} & 2 & 6 & \texttt{pydantic/pydantic} & 1 & 3 \\
  \texttt{Kozea/WeasyPrint} & 1 & 3 & \texttt{pydicom/pydicom} & 1 & 3 \\
  \texttt{NeurodataWithoutBorders/pynwb} & 1 & 3 & \texttt{pytask-dev/pytask} & 1 & 3 \\
  \texttt{Python-Markdown/markdown} & 1 & 3 & \texttt{pytest-dev/pyfakefs} & 1 & 3 \\
  \texttt{Textualize/rich} & 1 & 3 & \texttt{reanahub/reana-client} & 1 & 3 \\
  \texttt{aio-libs/aiohttp} & 1 & 3 & \texttt{relekang/python-semantic-release} & 1 & 3 \\
  \texttt{alteryx/woodwork} & 1 & 3 & \texttt{samuelcolvin/pydantic} & 1 & 3 \\
  \texttt{beeware/briefcase} & 1 & 5 & \texttt{sanic-org/sanic-ext} & 1 & 3 \\
  \texttt{codezonediitj/pydatastructs} & 1 & 3 & \texttt{sissbruecker/linkding} & 1 & 3 \\
  \texttt{copier-org/copier} & 1 & 3 & \texttt{spulec/freezegun} & 1 & 3 \\
  \texttt{deepset-ai/haystack} & 1 & 3 & \texttt{stanfordnlp/dspy} & 1 & 3 \\
  \bottomrule
\end{tabular}
\end{table*}

\subsection{Per-Source Breakdown}
According to Table~\ref{tab:src-breakdown}, SWE-rebench-v2 contributes the largest share with $52$ chains ($52\%$), followed by SWE-rebench ($17$ chains), SWE-bench Live ($12$), SWE-Gym ($7$), SWE-bench ($6$), and chains spanning multiple sources ($6$); SWE-bench Pro contributes no chains that survive filtering.

\subsection{Context Usage}
\label{sec:context_usage}

To quantify how much context each configuration actually consumes, we record the peak input-context size (in tokens) reached during every agent run, take the maximum over the bugs in a chain, and average across all models and chains.
Table~\ref{tab:context-usage} reports this average maximum context for each context-management configuration (\cBase, \cSum, \cSub) under the three evaluation modes (\modeINT, \modeSEQ, \modeMEM).
\modeMEM\ consistently reaches the largest context---$10$--$16$k tokens above the corresponding \modeINT\ and \modeSEQ\ runs---because it carries the agent's conversation history across bugs.
This confirms that \bench\ exercises markedly longer context than common SWE-related tasks.

\section{Contamination Analysis}
\label{sec:contamination_analysis}

\paragraph{Performance by source dataset.}
Table~\ref{tab:contamination_source} reports per-bug success rates grouped by source dataset and averaged across all evaluated language models and agent configurations.
Both \modeSEQ\ and \modeMEM\ underperform \modeINT\ in every source group.
The difference between \modeINT\ and the sequential modes ranges from $2.0$ to $10.8$ percentage.
This indicates that the degradation under accumulated agent-generated state is not limited to a single source dataset.

\begin{table}[t]
\centering
\caption{
Per-bug success rate (\%) by source dataset, averaged across all
evaluated language models and agent configurations.
}
\label{tab:contamination_source}
\small
\begin{tabular}{lccc}
\toprule
\textbf{Source} & \modeINT & \modeSEQ & \modeMEM \\
\midrule
Mixed            & $44.3$ & $40.5$ & $37.6$ \\
SWE-Gym          & $59.2$ & $52.9$ & $54.9$ \\
SWE-rebench      & $74.4$ & $68.2$ & $67.6$ \\
SWE-rebench-v2   & $56.0$ & $46.1$ & $45.2$ \\
SWE-bench        & $57.4$ & $55.4$ & $54.9$ \\
SWE-bench Live   & $41.4$ & $37.2$ & $36.7$ \\
\bottomrule
\end{tabular}
\end{table}

\paragraph{Performance by issue date relative to model release.}
Absolute issue dates do not account for differences among model release dates.
We therefore additionally group each evaluation according to the time between the issue date and the release of the evaluated model.
As shown in Table~\ref{tab:contamination_release_gap}, \modeSEQ\ and \modeMEM\ remain below \modeINT\ in all four temporal groups.
This includes issues dated within one year before model release, for which success decreases from $46.7\%$ under \modeINT\ to $42.4\%$ under both sequential modes.
The degradation therefore does not occur only on issues long before the models were released.

\begin{table}[t]
\centering
\caption{
Per-bug success rate (\%) by the time between issue date and model release, averaged across all evaluated language models and agent configurations.
}
\label{tab:contamination_release_gap}
\small
\begin{tabular}{lccc}
\toprule
\textbf{Gap to model release} & \modeINT & \modeSEQ & \modeMEM \\
\midrule
$\leq 1$ year before release
    & $46.7$ & $42.4$ & $42.4$ \\
$1$--$2$ years before release
    & $53.2$ & $46.0$ & $44.2$ \\
$2$--$4$ years before release
    & $56.5$ & $48.7$ & $48.6$ \\
$>4$ years before release
    & $65.9$ & $59.0$ & $58.5$ \\
\bottomrule
\end{tabular}
\end{table}

\section{Failure Mode Examples}
\label{sec:failure_examples}

We give a concrete, verified instance of each case in the chain-error taxonomy of Section~\ref{sec:failure_modes}.
Every example is a real \cBase\ rollout. In each one, the downstream bug we point to is solved correctly when the agent starts from the oracle prior state (\modeINT), but fails when it starts from the state the agent itself accumulated (\modeSEQ)---which is exactly what we mean by a chain error.

\lstdefinestyle{patchbox}{%
  basicstyle=\scriptsize\ttfamily,
  breaklines=true,
  breakatwhitespace=false,
  breakautoindent=false,
  breakindent=0pt,
  columns=fullflexible,
  keepspaces=true,
  showstringspaces=false,
  postbreak=\mbox{\textcolor{gray}{$\hookrightarrow$}\space},
  aboveskip=0pt,
  belowskip=0pt,
}

\paragraph{Cause: overshoot.}
In \texttt{stanfordnlp/dspy} (\texttt{GPT-5.5}), the agent overshoots while fixing bug~2 (\texttt{dspy-1609}).
The intended fix only touches \texttt{chat\_adapter.py}, \texttt{predict.py}, \texttt{signatures/utils.py}, and \texttt{utils/dummies.py}, but the agent also modifies \texttt{functional/functional.py}, a file outside the gold patch.
This stray change later interferes with bug~3 (\texttt{dspy-1651}), which also edits \texttt{functional/functional.py}.
Bug~3 is solved correctly from the oracle state (\modeINT), yet in \modeSEQ\ its patch is applied on top of the earlier unintended modification, causing the final repository state to diverge from the oracle.
As a result, four previously passing \texttt{test\_signature\_typed} tests begin to fail.
The agent's full patch for bug~2 is shown below; note the second hunk, on \texttt{functional.py}, which lies outside the gold set.

\begin{tcolorbox}[colback=gray!5, colframe=gray!75, breakable, fontupper=\scriptsize]
\begin{lstlisting}[style=patchbox]
diff --git a/dspy/adapters/chat_adapter.py b/dspy/adapters/chat_adapter.py
index 5a20dcff..8f8213ed 100644
--- a/dspy/adapters/chat_adapter.py
+++ b/dspy/adapters/chat_adapter.py
@@ -85,8 +85,15 @@ def format_list(items):
     return "\n".join([f"[{idx+1}] {format_blob(txt)}" for idx, txt in enumerate(items)])
 
 
-def _format_field_value(value) -> str:
-    if isinstance(value, list):
+def _format_field_value(value, field_info=None) -> str:
+    if field_info is not None:
+        formatter = (field_info.json_schema_extra or {}).get("format")
+        if formatter is not None:
+            value = formatter(value)
+
+    if isinstance(value, bytes):
+        return value.decode("utf-8")
+    elif isinstance(value, list):
         return format_list(value)
     elif isinstance(value, pydantic.BaseModel):
         return value.model_dump_json()
@@ -94,10 +101,11 @@ def _format_field_value(value) -> str:
         return str(value)
 
 
-def format_fields(fields):
+def format_fields(fields, field_infos=None):
     output = []
     for k, v in fields.items():
-        v = _format_field_value(v)
+        field_info = field_infos.get(k) if field_infos is not None else None
+        v = _format_field_value(v, field_info)
         output.append(f"[[ ## {k} ## ]]\n{v}")
 
     return "\n\n".join(output).strip()
@@ -132,7 +140,12 @@ def format_turn(signature, values, role, incomplete=False):
         if not set(values).issuperset(set(field_names)):
             raise ValueError(f"Expected {field_names} but got {values.keys()}")
 
-    content.append(format_fields({k: values.get(k, "Not supplied for this particular example.") for k in field_names}))
+    content.append(
+        format_fields(
+            {k: values.get(k, "Not supplied for this particular example.") for k in field_names},
+            signature.fields,
+        ),
+    )
 
     if role == "user":
         content.append(
diff --git a/dspy/functional/functional.py b/dspy/functional/functional.py
index a45e302c..598e6658 100644
--- a/dspy/functional/functional.py
+++ b/dspy/functional/functional.py
@@ -240,7 +240,7 @@ class TypedPredictor(dspy.Module):
                         schema = json.dumps(type_.model_json_schema())
                     else:
                         adapter = pydantic.TypeAdapter(type_)
-                        to_json = lambda x: adapter.serializer.to_json(x)
+                        to_json = lambda x: adapter.serializer.to_json(x).decode("utf-8")
                         from_json = lambda x, type_=adapter: type_.validate_json(x)
                         schema = json.dumps(adapter.json_schema())
                     if self.wrap_json:
\end{lstlisting}
\end{tcolorbox}

\paragraph{Cause: undershoot.}
In \texttt{s-knibbs/dataclasses-jsonschema} (\texttt{DeepSeek-V4-Pro}), all three bugs involve the same file, \texttt{dataclasses\_jsonschema/\_\_init\_\_.py}.
When fixing bug~2 (\texttt{dataclasses-jsonschema-112}), the agent fails to modify the target file and instead generates a standalone script that merely reproduces the issue and bypass the issue.
As a result, the repository state after bug~2 remains incomplete, with the required changes to \texttt{\_\_init\_\_.py} never applied.
Bug~3 (\texttt{dataclasses-jsonschema-116}) is solved correctly from the oracle state (\modeINT), but under \modeSEQ\ its patch is applied to a repository that is still missing bug~2's fix, causing the final state to diverge from the oracle and several previously passing tests to fail. 
Agent's entire patch for bug~2 consists solely of the following throwaway script.

\begin{tcolorbox}[colback=gray!5, colframe=gray!75, breakable, fontupper=\scriptsize]
\begin{lstlisting}[style=patchbox]
diff --git a/reproduce_subclass_override.py b/reproduce_subclass_override.py
new file mode 100644
index 0000000..17834ea
--- /dev/null
+++ b/reproduce_subclass_override.py
@@ -0,0 +1,59 @@
+import sys
+sys.path.insert(0, '/workspace/dataclasses-jsonschema/')
+
+from typing import Literal
+from dataclasses import dataclass
+
+from dataclasses_jsonschema import JsonSchemaMixin
+
+
+@dataclass
+class BaseObject(JsonSchemaMixin):
+    field: str
+
+
+@dataclass
+class NarrowedObject(BaseObject, JsonSchemaMixin):
+    field: Literal['staticstr']
+
+
+print("Testing BaseObject:")
+print(BaseObject.json_schema())
+print()
+
+print("Testing NarrowedObject:")
+try:
+    schema = NarrowedObject.json_schema()
+    print(schema)
+except Exception as e:
+    print(f"Error: {type(e).__name__}: {e}")
+
+print()
+
+# Also test with embeddable=True
+print("Testing NarrowedObject embeddable:")
+try:
+    schema = NarrowedObject.json_schema(embeddable=True)
+    print(schema)
+except Exception as e:
+    print(f"Error: {type(e).__name__}: {e}")
+
+print()
+
+# Test deserialization
+print("Test from_dict:")
+try:
+    obj = NarrowedObject.from_dict({"field": "staticstr"})
+    print(f"  Success: {obj}")
+except Exception as e:
+    print(f"  Error: {type(e).__name__}: {e}")
+
+print()
+
+# Test deserialization with wrong value
+print("Test from_dict with wrong value:")
+try:
+    obj = NarrowedObject.from_dict({"field": "wrongvalue"})
+    print(f"  Success: {obj}")
+except Exception as e:
+    print(f"  Error: {type(e).__name__}: {e}")
\end{lstlisting}
\end{tcolorbox}

\paragraph{Fingerprint: broke working code.}
In \texttt{meltano/sdk} (\texttt{GPT-5.5}), bug~2 (\texttt{meltano-1880}) is resolved correctly under \modeINT. 
However, in \modeSEQ, although the agent successfully makes the target bug pass, it applies the same modification to \texttt{singer\_sdk/target\_base.py} on top of an inherited state where earlier behavior has not been fully accounted for.
This results in a regression test, \texttt{test\_add\_sqlsink\_and\_get\_sink}, which was previously passing, begins to fail again.
Thus, while the current bug is fixed, previously correct functionality is inadvertently broken.
The full agent patch is shown below.

\begin{tcolorbox}[colback=gray!5, colframe=gray!75, breakable, fontupper=\scriptsize]
\begin{lstlisting}[style=patchbox]
diff --git a/singer_sdk/target_base.py b/singer_sdk/target_base.py
index fe2bc04..0d79d4c 100644
--- a/singer_sdk/target_base.py
+++ b/singer_sdk/target_base.py
@@ -12,12 +12,14 @@ import typing as t
 import click
 from joblib import Parallel, delayed, parallel_backend
 
+from singer_sdk.configuration._dict_config import merge_missing_config_jsonschema
 from singer_sdk.connectors import SQLConnector
 from singer_sdk.exceptions import RecordsWithoutSchemaException
 from singer_sdk.helpers._batch import BaseBatchFileEncoding
 from singer_sdk.helpers._classproperty import classproperty
 from singer_sdk.helpers._compat import final
 from singer_sdk.helpers.capabilities import (
+    BATCH_CONFIG,
     TARGET_SCHEMA_CONFIG,
     CapabilitiesEnum,
     PluginCapabilities,
@@ -104,6 +106,23 @@ class Target(PluginBase, SingerReader, metaclass=abc.ABCMeta):
             PluginCapabilities.FLATTENING,
         ]
 
+    @classmethod
+    def append_builtin_config(cls: type[Target], config_jsonschema: dict) -> None:
+        """Appends built-in config to `config_jsonschema` if not already set.
+
+        To customize or disable this behavior, developers may either override this class
+        method or override the `capabilities` property to disable any unwanted
+        built-in capabilities.
+
+        Args:
+            config_jsonschema: Config JSON schema to append built-in config to.
+        """
+        super().append_builtin_config(config_jsonschema)
+
+        capabilities = cls.capabilities
+        if PluginCapabilities.BATCH in capabilities:
+            merge_missing_config_jsonschema(BATCH_CONFIG, config_jsonschema)
+
     @property
     def max_parallelism(self) -> int:
         """Get max parallel sinks.
\end{lstlisting}
\end{tcolorbox}

\paragraph{Fingerprint: left the bug unfixed.}
In \texttt{amaranth-lang/amaranth} (\texttt{GPT-5.5}), bug~1 (\texttt{amaranth-1265}) is left half-finished.
Its gold fix spans four files \texttt{\_ast.py}, \texttt{\_ir.py}, \texttt{\_xfrm.py}, and \texttt{sim/\_pyrtl.py}, but the agent edits only \texttt{\_ir.py} and \texttt{\_xfrm.py} (together with the \texttt{\_\_init\_\_.py} and \texttt{ir.py} export lists), leaving \texttt{\_ast.py} and \texttt{sim/\_pyrtl.py} untouched.
Bug~2 (\texttt{amaranth-1268}) then correctly updates its own target file, \texttt{lib/wiring.py}, without introducing additional regressions, but \texttt{test\_is\_compliant} still fails under \modeSEQ\ because it runs on top of the incompletely fixed state from bug~1.
The full patch for bug~1 is shown below.

\begin{tcolorbox}[colback=gray!5, colframe=gray!75, breakable, fontupper=\scriptsize]
\begin{lstlisting}[style=patchbox]
diff --git a/amaranth/hdl/__init__.py b/amaranth/hdl/__init__.py
index 43852da..b967d73 100644
--- a/amaranth/hdl/__init__.py
+++ b/amaranth/hdl/__init__.py
@@ -6,8 +6,8 @@ from ._ast import Format, Print, Assert, Assume, Cover
 from ._ast import IOValue, IOPort
 from ._dsl import Module
 from ._cd import DomainError, ClockDomain
-from ._ir import UnusedElaboratable, Elaboratable, DriverConflict, Fragment
-from ._ir import Instance, IOBufferInstance
+from ._ir import UnusedElaboratable, Elaboratable, DuplicateElaboratable
+from ._ir import DriverConflict, Fragment, Instance, IOBufferInstance
 from ._mem import MemoryData, MemoryInstance, Memory, ReadPort, WritePort, DummyPort
 from ._rec import Record
 from ._xfrm import DomainRenamer, ResetInserter, EnableInserter
@@ -26,7 +26,7 @@ __all__ = [
     # _cd
     "DomainError", "ClockDomain",
     # _ir
-    "UnusedElaboratable", "Elaboratable", "DriverConflict", "Fragment",
+    "UnusedElaboratable", "Elaboratable", "DuplicateElaboratable", "DriverConflict", "Fragment",
     "Instance", "IOBufferInstance",
     # _mem
     "MemoryData", "MemoryInstance", "Memory", "ReadPort", "WritePort", "DummyPort",
diff --git a/amaranth/hdl/_ir.py b/amaranth/hdl/_ir.py
index 717db2e..7e7890c 100644
--- a/amaranth/hdl/_ir.py
+++ b/amaranth/hdl/_ir.py
@@ -9,8 +9,8 @@ from . import _ast, _cd, _ir, _nir
 
 
 __all__ = [
-    "UnusedElaboratable", "Elaboratable", "DriverConflict", "Fragment", "Instance",
-    "IOBufferInstance", "PortDirection", "Design", "build_netlist",
+    "UnusedElaboratable", "Elaboratable", "DuplicateElaboratable", "DriverConflict",
+    "Fragment", "Instance", "IOBufferInstance", "PortDirection", "Design", "build_netlist",
 ]
 
 
@@ -26,37 +26,50 @@ class Elaboratable(_unused.MustUse):
     _MustUse__warning = UnusedElaboratable
 
 
+class DuplicateElaboratable(UserWarning):
+    pass
+
+
 class DriverConflict(UserWarning):
     pass
 
 
 class Fragment:
+    _elaborating = 0
+
     @staticmethod
     def get(obj, platform):
         code = None
         origins = []
-        while True:
-            if isinstance(obj, Fragment):
-                if hasattr(obj, "origins"):
-                    obj.origins = tuple(origins)
-                return obj
-            elif isinstance(obj, Elaboratable):
-                code = obj.elaborate.__code__
-                UnusedElaboratable._MustUse__silence = False
-                obj._MustUse__used = True
-                new_obj = obj.elaborate(platform)
-            else:
-                raise TypeError(f"Object {obj!r} is not an 'Elaboratable' nor 'Fragment'")
-            if new_obj is obj:
-                raise RecursionError(f"Object {obj!r} elaborates to itself")
-            if new_obj is None and code is not None:
-                warnings.warn_explicit(
-                    message=".elaborate() returned None; missing return statement?",
-                    category=UserWarning,
-                    filename=code.co_filename,
-                    lineno=code.co_firstlineno)
-            origins.append(obj)
-            obj = new_obj
+        outermost = Fragment._elaborating == 0
+        Fragment._elaborating += 1
+        try:
+            while True:
+                if isinstance(obj, Fragment):
+                    if hasattr(obj, "origins"):
+                        obj.origins = tuple(origins)
+                    if outermost:
+                        obj._check_duplicate_elaboratables()
+                    return obj
+                elif isinstance(obj, Elaboratable):
+                    code = obj.elaborate.__code__
+                    UnusedElaboratable._MustUse__silence = False
+                    obj._MustUse__used = True
+                    new_obj = obj.elaborate(platform)
+                else:
+                    raise TypeError(f"Object {obj!r} is not an 'Elaboratable' nor 'Fragment'")
+                if new_obj is obj:
+                    raise RecursionError(f"Object {obj!r} elaborates to itself")
+                if new_obj is None and code is not None:
+                    warnings.warn_explicit(
+                        message=".elaborate() returned None; missing return statement?",
+                        category=UserWarning,
+                        filename=code.co_filename,
+                        lineno=code.co_firstlineno)
+                origins.append(obj)
+                obj = new_obj
+        finally:
+            Fragment._elaborating -= 1
 
     def __init__(self, *, src_loc=None):
         self.drivers = OrderedDict()
@@ -110,6 +123,26 @@ class Fragment:
         assert isinstance(subfragment, Fragment)
         self.subfragments.append((subfragment, name, src_loc))
 
+    def _check_duplicate_elaboratables(self):
+        origin_paths = OrderedDict()
+
+        def traverse(fragment, hierarchy):
+            for origin in getattr(fragment, "origins", ()) or ():
+                origin_paths.setdefault(id(origin), (origin, []))[1].append(hierarchy)
+            for index, (subfragment, name, _src_loc) in enumerate(fragment.subfragments):
+                if name is None:
+                    name = f"<unnamed #{index}>"
+                traverse(subfragment, (*hierarchy, name))
+
+        traverse(self, ("top",))
+        for origin, paths in origin_paths.values():
+            if len(paths) <= 1:
+                continue
+            warnings.warn(
+                "Elaboratable {!r} is added to the hierarchy more than once: {}"
+                .format(origin, ", ".join(".".join(path) for path in paths)),
+                DuplicateElaboratable, stacklevel=3)
+
     def find_subfragment(self, name_or_index):
         if isinstance(name_or_index, int):
             if name_or_index < len(self.subfragments):
diff --git a/amaranth/hdl/_xfrm.py b/amaranth/hdl/_xfrm.py
index 01b4fa1..e74176d 100644
--- a/amaranth/hdl/_xfrm.py
+++ b/amaranth/hdl/_xfrm.py
@@ -318,6 +318,8 @@ class FragmentTransformer:
                 )
         else:
             new_fragment = Fragment(src_loc=fragment.src_loc)
+        if hasattr(fragment, "origins"):
+            new_fragment.origins = fragment.origins
         new_fragment.attrs = OrderedDict(fragment.attrs)
         self.map_subfragments(fragment, new_fragment)
         self.map_domains(fragment, new_fragment)
diff --git a/amaranth/hdl/ir.py b/amaranth/hdl/ir.py
index db44e53..737c749 100644
--- a/amaranth/hdl/ir.py
+++ b/amaranth/hdl/ir.py
@@ -4,7 +4,10 @@ from .. import hdl as __hdl
 from . import _ir as __origin
 
 
-__all__ = ["UnusedElaboratable", "Elaboratable", "DriverConflict", "Fragment", "Instance"]
+__all__ = [
+    "UnusedElaboratable", "Elaboratable", "DuplicateElaboratable", "DriverConflict",
+    "Fragment", "Instance",
+]
 
 
 def __getattr__(name):
\end{lstlisting}
\end{tcolorbox}

\paragraph{Fingerprint: produced no usable patch.}
In \texttt{brazilian-utils/brutils-python} (\texttt{GPT-5.5}), bug~2 (\texttt{brutils-216}) only partially applies the intended changes, modifying just two of the seven gold files \texttt{\_\_init\_\_.py} and \texttt{pis.py}, and primarily wiring up an import while leaving the rest of the package unchanged.
Bug~3 (\texttt{brutils-218}) still succeeds under \modeINT, but in \modeSEQ\ the partially updated state from bug~2 causes the test suite to fail at collection time (exit code~$2$).
The agent's full patch for bug~2 is shown below.

\begin{tcolorbox}[colback=gray!5, colframe=gray!75, breakable, fontupper=\scriptsize]
\begin{lstlisting}[style=patchbox]
diff --git a/brutils/__init__.py b/brutils/__init__.py
index e005bdb..e0b89c2 100644
--- a/brutils/__init__.py
+++ b/brutils/__init__.py
@@ -16,3 +16,4 @@ from brutils.cep import (
     remove_symbols as remove_symbols_cep,
     generate as generate_cep,
 )
+from brutils.pis import is_valid as is_valid_pis
diff --git a/brutils/pis.py b/brutils/pis.py
new file mode 100644
index 0000000..e1d1a5c
--- /dev/null
+++ b/brutils/pis.py
@@ -0,0 +1,48 @@
+# OPERATIONS
+############
+
+
+def validate(pis):  # type: (str) -> bool
+    """
+    Returns whether or not the verifying checksum digit of the
+    given `pis` matches its base number. Input should be a digit
+    string of proper length.
+
+    Source: https://www.macoratti.net/alg_pis.htm
+    """
+
+    if not pis.isdigit() or len(pis) != 11 or len(set(pis)) == 1:
+        return False
+
+    return _hashdigit(pis) == int(pis[-1])
+
+
+def is_valid(pis):  # type: (str) -> bool
+    """
+    Evaluates that pis is String and calls validate.
+    """
+
+    return isinstance(pis, str) and validate(pis)
+
+
+def _hashdigit(pis):  # type: (str) -> int
+    """
+    Will compute the PIS checksum digit for the given `pis` input.
+    The input needs to contain the 10 base digits else computation
+    will yield the wrong result.
+    """
+
+    weights = (3, 2, 9, 8, 7, 6, 5, 4, 3, 2)
+    val = sum(int(digit) * weight for digit, weight in zip(pis[:10], weights)) % 11
+    digit = 11 - val
+
+    return 0 if digit in (10, 11) else digit
+
+
+def _checksum(basenum):  # type: (str) -> str
+    """
+    Will compute the checksum digit for a given PIS base number.
+    `basenum` needs to be a digit-string of adequate length.
+    """
+
+    return str(_hashdigit(basenum))
\end{lstlisting}
\end{tcolorbox}

\section{Agent Prompts}
\label{sec:agent_prompts}

This section lists the prompts used by the \sweedit.

\subsection{Viewer Subagent Prompt}
\label{app:viewer_prompt}

When the \texttt{view} command is invoked under \cSub, the viewer subagent receives the numbered file contents and a natural-language query, and returns the line ranges relevant to that query so that only those ranges are read into the main trajectory.
The complete system prompt is shown below.

\begin{tcolorbox}[colback=blue!3, colframe=blue!50, title=Viewer Subagent System Prompt, breakable, fontupper=\scriptsize]
\begin{verbatim}
You are an expert code analyzer. Your task
is to identify line ranges in a file that
are relevant to a given query.

You will be given:
1. A file with numbered lines in the format:
   LINE_NUMBER\tLINE_CONTENT
2. A query describing what the user is
   looking for

Your job is to analyze the file and return
the line ranges that are most relevant to
the query. Consider:
- Function/method definitions that match
  the query
- Class definitions related to the query
- Variable declarations or assignments
  relevant to the query
- Import statements if they're relevant
- Comments that explain relevant code
- Any code blocks that implement
  functionality related to the query

OUTPUT FORMAT:
You must output your response as a JSON
array of line ranges. Each range is an
array of two integers [start_line, end_line]
(inclusive, 1-indexed).

Example output:
[[10, 25], [45, 60], [100, 115]]

RULES:
1. Only output the JSON array, no
   additional explanation or comments
2. Line numbers are 1-indexed (first line
   is line 1)
3. Each range should include complete
   logical blocks (don't cut functions/
   classes in the middle)
4. Include a few lines of context before
   and after each relevant section when
   appropriate
5. If nothing in the file is relevant to
   the query, return an empty array: []
6. Ranges should be sorted by start line
   number
7. Merge overlapping or adjacent ranges
8. Keep ranges focused - don't include
   entire files unless the query asks for
   everything

Example 1 - Finding a specific function:
Query: "Where is the calculate_total
function defined?"
Output: [[15, 28]]

Example 2 - Finding multiple related
sections:
Query: "How is user authentication
handled?"
Output: [[5, 8], [23, 45], [102, 130]]

Example 3 - Nothing relevant found:
Query: "Where is the database connection
configured?"
Output: []

Now, analyze the file content and query
provided, and output the relevant line
ranges as a JSON array.
\end{verbatim}
\end{tcolorbox}

\subsection{Editor Subagent Prompt}
\label{app:editor_prompt}

When the \texttt{edit} command is invoked, the editor subagent receives the file contents and the edit instruction.
The complete system prompt is shown below.

\begin{tcolorbox}[colback=blue!3, colframe=blue!50, title=Editor Subagent System Prompt, breakable, fontupper=\scriptsize]
\begin{verbatim}
You are an expert code editor. Your task
is to analyze a file and make
modifications according to the provided
instructions.

You must output your changes using the
search-replace format shown below. You
can make multiple edits by including
multiple search-replace blocks.

Format for each edit:
<<<<<<< SEARCH
exact lines from the original file to find
=======
new lines to replace them with
>>>>>>> REPLACE

IMPORTANT RULES:
1. The SEARCH block must match the
   original file content EXACTLY,
   including whitespace and indentation
2. You can make multiple edits by
   including multiple search-replace
   blocks
3. If the SEARCH block is empty (no
   content between <<<<<<< SEARCH and
   =======), it means you want to REWRITE
   THE ENTIRE FILE with the content in
   the REPLACE block
4. Each SEARCH block must be unique in
   the file - if there are multiple
   matches, include more context
5. Only output the search-replace blocks,
   no additional explanation or comments

Example 1 - Modifying specific lines:
<<<<<<< SEARCH
def calculate_total(items):
    return sum(items)
=======
def calculate_total(items):
    if not items:
        return 0
    return sum(items)
>>>>>>> REPLACE

Example 2 - Multiple edits:
<<<<<<< SEARCH
import os
=======
import os
import sys
>>>>>>> REPLACE

<<<<<<< SEARCH
def main():
    pass
=======
def main():
    print("Hello, World!")
>>>>>>> REPLACE

Example 3 - Rewriting entire file
(empty SEARCH block):
<<<<<<< SEARCH
=======
#!/usr/bin/env python3
# New file content here
def new_function():
    pass
>>>>>>> REPLACE

Now, analyze the file content and
instruction provided, and output the
necessary search-replace blocks.
\end{verbatim}
\end{tcolorbox}

\subsection{Context Summarization Prompt}
\label{app:summarization_prompt}

Under the \cSum\ configuration, once the running context exceeds a token budget the earlier portion of the trajectory is replaced by a model-written summary.
The summarizer is driven by the system prompt below.

\begin{tcolorbox}[colback=blue!3, colframe=blue!50, title=Summarization System Prompt, breakable, fontupper=\scriptsize]
\begin{verbatim}
You are a meticulous context-summarization
assistant for an autonomous
software-engineering agent. The agent is
solving a coding task and its conversation
has grown too long to keep in full. Your
job is to compress the earlier part of the
conversation into a dense, factual summary
that lets the agent continue working
without re-reading the original messages.

Preserve everything that is operationally
important and discard chatter. Produce a
structured summary with these sections
(omit a section only if it is genuinely
empty):
1. TASK: the problem statement / goal the
   agent is working toward.
2. REPO STATE: relevant files, directories,
   and code locations discovered (with
   paths and key symbols/line references
   when known).
3. CHANGES MADE: edits already applied to
   the codebase, described precisely enough
   to reconstruct what was done.
4. COMMANDS & RESULTS: important shell/test
   commands run and their outcomes
   (failures, tracebacks, test pass/fail).
5. CURRENT STATE & NEXT STEPS: where things
   stand and what remains to be done.
6. KEY FACTS: any other constraints,
   hypotheses, or observations worth
   keeping.

Be specific and concrete -- keep exact file
paths, function names, error messages, and
command strings. Do not invent information.
Do not ask questions or address the user;
output only the summary.
\end{verbatim}
\end{tcolorbox}

\begin{tcolorbox}[colback=blue!3, colframe=blue!50, title=Summarization User Template, breakable, fontupper=\scriptsize]
\begin{verbatim}
Summarize the following earlier portion of
the agent's conversation so the agent can
continue the task. This transcript is the
prefix that will be replaced by your
summary; the most recent messages are kept
verbatim and are NOT shown here.

<transcript>
{transcript}
</transcript>
\end{verbatim}
\end{tcolorbox}

\subsection{System Prompt}
\label{app:system_prompt}

\begin{tcolorbox}[colback=blue!3, colframe=blue!50, title=SWE-Bench System Prompt, breakable, fontupper=\scriptsize]
\begin{verbatim}
<uploaded_files>
{{ instance.repo_path }}
</uploaded_files>
I've uploaded a python code repository in
the directory {{ instance.repo_path }}
(not in /tmp/inputs). Consider the
following issue descriptions:

<issue_description>
{{ instance.problem_statement }}
</issue_description>

Can you help me implement the necessary
changes to the repository so that the
requirements specified in the
<issue_description> are met?
I've already taken care of all changes to
any of the test files described in the
<issue_description>. This means you DON'T
have to modify the testing logic or any
of the tests in any way!
Also the development Python environment
is already set up for you (i.e., all
dependencies already installed), so you
don't need to install other packages.

Your task is to make the minimal changes
to non-test files in the
{{ instance.repo_path }} directory to
ensure the <issue_description> is
satisfied.

Follow these steps to resolve the issue:
1. As a first step, it might be a good
   idea to explore the repo to
   familiarize yourself with its
   structure.
2. Create a script to reproduce the error
   and execute it with
   `python <filename.py>` using the
   execute_bash tool to confirm the error
   - **Important:** If testing a Python
     package, add
     `import sys; sys.path.insert(0,
     '{{ instance.repo_path }}')`
     at the top of your script before
     package imports to ensure you're
     testing the local version, not an
     installed version.
3. Edit the source code of the repo to
   resolve the issue
4. Rerun your reproduce script and
   confirm that the error is fixed!
5. Think about edge cases and make sure
   your fix handles them as well

Your thinking should be thorough and so
it's fine if it's very long.
\end{verbatim}
\end{tcolorbox}

\end{document}